\documentclass[conference,hidelinks]{IEEEtran}

\usepackage{cite}
\usepackage{amsmath,amssymb,amsfonts}
\usepackage{graphicx}
\usepackage{xcolor}
\usepackage[ruled,vlined]{algorithm2e}
\usepackage{booktabs}
\usepackage{algpseudocode}
\usepackage{orcidlink}
\title{Low-Complexity Sensing-Aware PAPR Reduction for AFDM-based ISAC Systems }

\author{
\IEEEauthorblockA{
Eya Gourar\IEEEauthorrefmark{1}, Abdul Karim Gizzini\IEEEauthorrefmark{2}, Yahia Medjahdi\IEEEauthorrefmark{1}, Patrick Sondi\IEEEauthorrefmark{1}, Laurent Clavier\IEEEauthorrefmark{3}
}\\
\IEEEauthorblockA{
\IEEEauthorrefmark{1} IMT Nord Europe, Institut Mines T\'el\'ecom, Center for Digital Systems, F-59653 Villeneuve d’Ascq, France\\
\IEEEauthorrefmark{2} University of Paris-Est Créteil (UPEC), LISSI/TincNET, F-94400, Vitry-sur-Seine, France\\
\IEEEauthorrefmark{3} Inria, Villeneuve-d’Ascq, France
}
}

\begin{document}
\maketitle

\begin{abstract}
Integrated sensing and communication (ISAC) has emerged as a key technology for future wireless networks by enabling communication and environmental sensing through a common waveform and hardware platform. Among the candidate waveforms for ISAC, Affine Frequency Division Multiplexing (AFDM) had attracted significant attention due to its robustness in high-mobility environments, but it suffers from a high peak-to-average power ratio (PAPR). In this paper, we propose a sensing-aware chirp-subcarrier reservation (CSR) framework that reduces PAPR while improving ranging performance. The proposed method combines low-complexity gradient-based PAPR minimization with a randomized local search that exploits the phase sensitivity of the AFDM autocorrelation function to suppress delay low-ambiguity-zone (LAZ) sidelobes. Numerical results show that the proposed scheme achieves significant PAPR reduction together with significant sidelobe suppression, resulting in improved weak-target detection performance.
\end{abstract}


\section{Introduction}
Integrated sensing and communication (ISAC) has emerged as a fundamental enabling technology for future sixth-generation (6G) wireless networks by allowing sensing and communication functionalities to share the same spectrum resources, hardware infrastructure, and signal processing chain \cite{liu2022survey}. The sensing performance of a waveform is commonly characterized through its ambiguity function (AF), which describes the delay-Doppler response of the sensing matched filter at the receiver \cite{liu2022survey}. In particular, the AF sidelobes quantify the level of interference caused by neighboring targets, making them a critical metric for multi-target sensing in 6G systems. It has been reported that Orthogonal Frequency Division Multiplexing (OFDM) exhibits the lowest ranging sidelobes, \textit{i.e.}, delay-domain AF sidelobes, resulting in remarkable sensing performance~\cite{liu2025cp}. Consequently, OFDM has served as the baseline waveform during the early development of ISAC. However, its sensitivity to Doppler spread introduces inter-carrier interference, thereby degrading both communication reliability and sensing accuracy in high-mobility environments \cite{wang2006performance}. To overcome these limitations, Affine Frequency Division Multiplexing (AFDM) has recently been introduced as a flexible multicarrier waveform that leverages the delay-Doppler characteristics of doubly selective channels \cite{bemani2021afdm}. As a result, the sensing performance of AFDM has attracted considerable research attention. A growing body of literature has investigated the AF characteristics of AFDM waveforms (see, \textit{e.g.}, \cite{rou2025normalized,zhang2026afdmenabled}), demonstrating several notable advantages. Nevertheless, AFDM has been shown to inherit a practical limitation also encountered in OFDM systems, namely its high Peak-to-Average Power Ratio (PAPR) \cite{bemani2021afdm}, which pushes the power amplifier into its nonlinear operating region, resulting in signal distortion and degraded system performance. Therefore, efficient PAPR reduction remains a key requirement for practical AFDM implementations. Existing PAPR reduction techniques for AFDM are relatively scarce and include $\rm DAFT$-spread transmission architectures \cite{tao2026affine} as well as grouped pre-chirp selection schemes \cite{yuan2025papr}, the latter requiring additional side information to be transmitted to the receiver. More recently, a unified AFDM waveform design framework based on the optimization of Reserved Chirp-Subcarrier (RCS) symbols and, optionally, pre-chirp parameters using a majorization-minimization approach was proposed in \cite{meng2026afdm_papr_isl}. While this framework offers substantial flexibility in waveform design, its computational complexity becomes significant even when a relatively small number of RCSs (20\% of the bandwidth) is employed, and only moderate sensing improvements are sought. Moreover, the receiver's demodulator must be updated accordingly about the selected pre-chirp parameters, in a bistatic scenario.

Motivated by these observations, this paper investigates a different waveform design perspective. Specifically, we build on the observation that AFDM with random signaling exhibits Doppler-domain characteristics comparable to those of OFDM while suffering from significantly higher ranging sidelobes \cite{rou2025normalized}. Rather than pursuing a full AF optimization, we pursue the practical objective of reducing the PAPR of the AFDM waveform while ensuring acceptable ranging performance. Hence, we exploit RCS as waveform design degrees of freedom without altering the underlying AFDM structure. The main contributions of this paper are summarized as follows:
\begin{itemize}
\item We propose a lightweight sensing-aware Chirp-Subcarrier Reservation (CSR) framework that combines gradient-based PAPR minimization with a randomized local-search procedure to reduce ranging sidelobes over a prescribed delay low-ambiguity zone (LAZ).
\item Through numerical simulations, we demonstrate significant PAPR reduction and near-mainlobe sidelobe suppression, yielding improved weak-target detection performance.
\item We analyze the computational complexity of the proposed algorithm and show that it achieves sensing performance comparable to the state-of-the-art while requiring fewer operations.
\end{itemize}

The remainder of this paper is organized as follows: Section II introduces the AFDM signal model and formulates the PAPR reduction problem. Section III presents the proposed sensing-aware CSR framework and its computational complexity. Section IV provides numerical results demonstrating the performance of the proposed method. Finally, Section V concludes the paper.

\section{Preleminaries}
Consider an AFDM-based ISAC system where a single DAFT-domain symbol $
\boldsymbol{z} = [z_0, z_1, \ldots, z_{N-1}]^T \in \mathbb{C}^{N \times 1}$ is transmitted over $N$ chirp subcarriers. 
 The discrete-time AFDM transmit waveform is obtained via the inverse $\rm DAFT$ ($\rm IDAFT$),
\begin{equation}
\mathbf{x} =
\mathbf{A}^H \boldsymbol{z} =
\mathbf{\Lambda}_{c_1}^H
\mathbf{F}^H
\mathbf{\Lambda}_{c_2}^H
\boldsymbol{z}
\in \mathbb{C}^{N \times 1}
\end{equation}
where $\mathbf{F}$ denotes the $N$-point DFT matrix and $\mathbf{\Lambda}_{c_1}$, $\mathbf{\Lambda}_{c_2}$ are diagonal chirp matrices, defined by $\boldsymbol{\Lambda}_{c_i} =\mathrm{diag}\left(1,e^{-j2\pi c_i},\dots,e^{-j2\pi c_iN^2}\right)$ with a central digital frequency of $c_i$.

The integrated transmitted signal $\mathbf{x}$ can be optimized to compromise between the communication performance and sensing performance, by partitioning the index set $ \mathcal{N} = \{0,1,\dots,N-1\} $ into two disjoint subsets; the data-carrying chirp-subcarriers $\mathcal{D}$ and the RCS $\mathcal{R}$, such that $\mathcal{N} = \mathcal{D} \cup \mathcal{R}$ and $\mathcal{D} \cap \mathcal{R} = \varnothing$, with $|\mathcal{D}| = N_c$, $|\mathcal{R}| = N_r$, and $N = N_c + N_r$. We define the Data-to-Bandwidth Ratio $\mathrm{DBR}= \frac{N_c}{N} = 1-\frac{N_r}{N}$, which accounts for the spectral efficiency.

The data sequence $\boldsymbol{d} = \Big\{ \{\boldsymbol{z}_m\}_{m \in \mathcal{D}} \cup \mathbf{0}_{N_r \times 1}\Big\}$ is drawn from a random symbols constellation, while the RCS $\boldsymbol{r} \in \mathbb{C}^{N_r \times 1}$ can be optimized directly as waveform-design variables. By separating the contributions of data and reserved modes, the TD signal can be written as,
\begin{align}
\mathbf{x} &= \mathbf{A}^H \boldsymbol{d} + \mathbf{A}_{\mathcal{R}}^H \boldsymbol{r} 
= \mathbf{d}+ \mathbf{A}_{\mathcal{R}}^H \boldsymbol{r},
\label{eq:x_final}
\end{align} 
where $\mathbf{d} \in \mathbb{C}^{N \times 1}$ is the time-domain (TD) data sequence and $\mathbf{A}_{\mathcal{R}}^H \in \mathbb{C}^{N\times N_r}$ is a modified $\rm IDAFT$ matrix with each column corresponding to the position of a sensing pilot in the AFDM signal, allowing reduced modulation complexity \cite{rexhepi2025tone}.

A critical metric in ISAC systems is the peak-to-average power ratio (PAPR), defined as,
\begin{equation}
\mathrm{PAPR}(\mathbf{x})
\frac{\|\mathbf{x}\|_\infty^2}
{\frac{1}{N}\|\mathbf{x}\|_2^2}.
\end{equation}
where $\|\mathbf{x}\|_\infty$ and $\|\mathbf{x}\|_2$ denote the $\ell_\infty$ and $\ell_2$ norm operators, respectively. PAPR is a significant drawback in multicarrier systems due to the superposition of many subcarriers producing high instantaneous peaks, which causes nonlinear distortion in power amplifiers and degrades both communication reliability and sensing accuracy.
\section{Problem Formulation And Proposed Solution}
The PAPR minimization can be considered as the problem of finding the optimal RCS $\boldsymbol{r}$, which we refer to as the Chirp-Subcarrier Reservation (CSR), in this paper.

\subsection{Reformulating Existing Technology: Gradient-descent Chirp-Subcarrier Reservation}
Leveraging the reformulated notation in \eqref{eq:x_final}, and noticing that minimizing $\|x\|_p$ also minimizes $\|x\|_p^2$, the CSR can be formulated to reduce processing complexity as,
\begin{align}
\mathcal P_1:&\quad \arg \min_{\boldsymbol{r} \in \mathbb{C}^{N_r}}
\|\mathbf{x}\|_\infty,\\
& \qquad\text {s.t.} \quad \|\boldsymbol{r}\|^2 \leq P_{\max} \;,
\end{align}
where $P_{\max}$ is a power constraint. This PAPR minimization framework can be conducted with a gradient-descent-based technique \cite{rexhepi2025tone}. Because the $\ell_\infty$ norm is non-differentiable, it is approximated using the $\ell_p$ norm for sufficiently large $p$ as
\begin{equation}
\|\mathbf{x}\|_\infty \approx \lim_{p \rightarrow \infty} \|\mathbf{x}\|_p = \left(
\sum_{n=0}^{N-1} |x_n|^p
\right)^{1/p}.
\end{equation}
The gradient of $p$-norm with respect to $\boldsymbol{r}$ is calculated with the following \cite{rexhepi2025tone},
\begin{equation}
\nabla_{\boldsymbol{r}} f = \frac{1}{\|\mathbf{x}\|_p^{p-1}}
\mathbf{A}_{\mathcal{R}} (\mathbf{
|\mathbf{x}|^{p-2} \odot \mathbf{x}})
\label{eq:classical_gradient}
\end{equation}
where $\odot$ denotes element-wise multiplication.

We update each iteration $k$ of the gradient descent with the following expression,
\begin{equation}
\boldsymbol{r}^{(k+1)} = 
\boldsymbol{r}^{(k)}
-
\alpha
\nabla_{\boldsymbol{r}} f
\big(
\boldsymbol{r}^{(k)}
\big),
\label{eq:r_k+1}
\end{equation}
where $\alpha > 0$ is the step size. This constitutes the gradient-descent-based CSR approach for AFDM, which directly extends the OFDM Tone Reservation (TR) framework (see e.g, \cite{rexhepi2025tone}) into the $\rm DAFT$ domain.

\subsection{Bi-objective optimization problem}
At the receiver of the ISAC system, multiple echoes of the transmitted signal $\mathbf{x}$ are received after being reflected by the targets. The radar image is then reconstructed using the matched filtering technique, which involves cross-correlating the received signal with $\mathbf{x}$. Consequently, the quality of the matched filter response depends on the autocorrelation function (ACF) of $\mathbf{x}$, which is defined as follows,
\begin{equation}
\mathcal A(l)
=
\sum_{n=0}^{N-1}
x_nx^{*}_{n-l},
\quad l=0,\ldots,N-1,
\label{eq:ACF}
\end{equation}
where $(\cdot)^*$ denotes the complex conjugate. A commonly used metric measures the energy of the ACF sidelobes and is known as the integrated sidelobe level (ISL). Ideally, we would like ISL to be optimally low, to resolve highly correlated targets. We define $\mathcal W$ as the set of delay indices outside the mainlobe support. The sensing metric is then defined as,
\begin{equation}
J(\boldsymbol{r}) = \sum_{l\in\mathcal W}
\left|
\mathcal A(l)
\right|^2.
\end{equation}
We can see that ISAC problems are inherently multi-objective optimization problems. Hence, the problem at hand can be expressed as a bi-objective optimization,
\begin{align}
\mathcal P_2:&\quad \min_{\boldsymbol r}
\Big(
f_{\mathrm{PAPR}}(\boldsymbol r),
J(\boldsymbol r)
\Big)\\
\mathrm{s.t.}
&\quad
|r_i|=1,\qquad \forall i\in\mathcal R,
\end{align}
where $ f_{\mathrm{PAPR}}(\boldsymbol r) $ is the smooth approximation of the $\ell_\infty$ norm introduced in the previous subsection, and the unimodular constraint limits the power allocated to the RCS, ensuring that the total reserved-chirp energy remains fixed and equal to $N_r$. However, obtaining a globally optimal solution may become computationally heavy \cite{vaezi2026tutorial}. In OFDM, simply normalizing $\boldsymbol r$ after the update rule in \eqref{eq:r_k+1} (while fixing $c_1\!=\!c_2\!=\!0$), at each iteration ensures that the reserved tones are located on the unit complex circle, enabling theoretically perfect detection properties, which is not similar for AFDM \cite{liu2025cp}. Therefore, we need to highlight how the RCS provide useful degrees of freedom for sidelobe shaping in AFDM.
\subsection{Autocorrelation Sensitivity to the RCS} \label{sec:sensitivity}
Indeed, the OFDM ACF collapses to a function of $|z_m|^2$ under the unimodularity constraint. Therefore, its sidelobe structure is primarily controlled by the power distribution over subcarriers.
However, for the AFDM waveform, the ACF is expressed differently;\cite{zhang2026afdmenabled}
\begin{align}
\mathcal A(l)
&=
\frac{1}{N}
\sum_{n=0}^{N-1}
\sum_{m_1=0}^{N-1}
\sum_{m_2=0}^{N-1}
z_{m_1}z^*_{m_2}
e^{j2\pi\left(c_1 n^2 + \frac{m_1 n}{N} + c_2 m_1^2\right)} \nonumber
\\
&\hspace{2cm}\times
e^{-j2\pi\left(c_1 (n-l)_N^2 + \frac{m_2 (n-l)_N}{N} + c_2 m_2^2\right)} \nonumber
\\
&=
\frac{1}{N}
\sum_{m_1=0}^{N-1}
\sum_{m_2=0}^{N-1}
z_{m_1}z^*_{m_2},
S_{l}^{(c_1,c_2)}(m_1,m_2),
\label{eq:afdm_acf}
\end{align}
where
\begin{align*}
S_{l}^{(c_1,c_2)}(m_1,m_2)
&=
\sum_{n=0}^{N-1}
e^{j2\pi
c_1\left(n^2-(n-l)_N^2\right)} \nonumber \\
& \qquad \quad \times e^{j2\pi
\frac{m_1 n-m_2(n-l)_N}{N}
+
c_2(m_1^2-m_2^2)} .
\end{align*}
Hence, unlike OFDM, the AFDM ACF depends not only on the terms $|z_m|^2$, but also on the cross-terms $z_{m_1}z_{m_2}^{*}$ for $m_1\neq m_2$, weighted by the chirp-dependent coefficients $S_{l}^{(c_1,c_2)}(m_1,m_2)$. Consequently, the sidelobe structure is not determined solely by the subcarrier power distribution, but also by the relative phases between DAFT-domain symbols and by the chirp parameters $c_1$ and $c_2$. Therefore, even small changes in the RCS may alter the phase alignment among the cross-terms, which can either reinforce or attenuate the existing sidelobes through constructive or destructive interference, respectively. To characterize this effect, we denote the reserved chirp vector obtained after the $k^\text{th}$ PAPR gradient update as,
\begin{align}
\tilde{\boldsymbol r}^{(k+1)}
=
\boldsymbol r^{(k)}
-
\alpha
\nabla_{\boldsymbol r} f(\boldsymbol r^{(k)}).
\end{align}
From a random additive perturbation $\boldsymbol u$, we form two candidates as,
\begin{align}
\boldsymbol r_{\pm}
=
\tilde{\boldsymbol r}^{(k+1)}
\pm
\delta\boldsymbol u,
\qquad
\boldsymbol u\in\mathbb C^{N_r}.
\end{align}
where $\delta >0$ is the ISL control coefficient. The corresponding full DAFT-domain reserved vector is,
\begin{align}
y_{\pm,m}
=
\begin{cases}
r_{\pm,m}, & m\in\mathcal R,\\
0, & m\notin\mathcal R.
\end{cases}
\end{align}

The modified ACF can be written as,
\begin{align}
\mathcal A_{\pm}(l)
=
\mathcal A_d(l)
+
\mathcal A_{r_{\pm}}(l)
+
\mathcal A_{d r_{\pm}}(l)
+
\mathcal A_{r_{\pm} d}(l),
\label{eq:acf_rcs_decomposition_perturbed}
\end{align}
where the self-ACF and cross-ACF are defined, respectively, as,
\begin{align*}
    \mathcal{A}_{w}(l) = \sum_{m_1\in\mathcal D}
\sum_{m_2\in\mathcal D}
w_{m_1}w_{m_2}^{*}
S_{l}^{(c_1,c_2)}(m_1,m_2), \\
    \mathcal{A}_{w,g}(l) =  \sum_{m_1\in\mathcal D}
\sum_{m_2\in\mathcal R}
w_{m_1} g_{m_2}^{*}
S_{l}^{(c_1,c_2)}(m_1,m_2).
\end{align*}

Since the perturbation is applied only to the RCS, we write, for $m\in\mathcal R$, $y_{\pm,m} =
\tilde y_{m}^{(k+1)}
\pm
\delta u_{i(m)}$, where $i(m)$ denotes the index of the RCS $m$ in $\boldsymbol r$. Equation \eqref{eq:acf_rcs_decomposition_perturbed} becomes
\begin{align}
\mathcal A_{\pm}(l)
=
\mathcal A_{\tilde r}(l)
\pm
\delta
B(l)
+
\delta^{2}
C(l),
\label{eq:rcs_perturbed_acf_compact}
\end{align}
where $\mathcal A_{\tilde r}(l)$ denotes the AFDM autocorrelation obtained after the PAPR gradient update, namely
\begin{align*}
\mathcal A_{\tilde r}(l)
&=
\mathcal A_d(l)
+
\mathcal A_{\tilde r}(l)
+
\mathcal A_{d\tilde r}(l)
+
\mathcal A_{\tilde r d}(l),
\end{align*}
the first-order RCS-induced perturbation term is given by
\begin{align*}
B(l)
&= \mathcal{A}_{u,\tilde y^{(k+1)}} (l)+ \mathcal{A}_{\tilde y^{(k+1)},u}(l) + \mathcal{A}_{d,u}(l) + \mathcal{A}_{u,d}(l),
\end{align*}
and the second-order term is $C(l)= \mathcal{A}_{u,u}$.

Equation \eqref{eq:rcs_perturbed_acf_compact} shows that the two candidates $\boldsymbol r_{+}$ and $\boldsymbol r_{-}$ can generate opposite impacts in the ACF depending on the ISL-control parameter $\delta$, making the RCS perturbation a controllable mechanism. As $\delta$ increases, $\pm \delta B(l)$ becomes more significant, enhancing the impact of the RCS on the sidelobes. This growing sensitivity enables a wider RCS search space and may lead to solutions with lower ISL. However, for large values of $\delta$, the sidelobe structure becomes governed by $\delta^2 C(l)$, rather than the optimized solutions obtained from the preceding PAPR minimization, which can become significantly distorted. Therefore, $\delta$ should be fine-tuned accordingly.

\subsection{Proposed Sensing-Aware Chirp-Subcarrier Reservation}
In practice, the maximum observable delay $l_\text{max}$ is significantly smaller than the signal duration \cite{meng2026afdm_papr_isl}. Consequently, the sidelobes located in the vicinity of the mainlobe, denoted by the delay low ambiguity zone (LAZ), are of primary importance, given that they directly determine the ability to detect weak targets near the strong reflectors.

To solve $\mathcal P_2$, we propose an iterative sequential optimization method based on the study in subsection~\ref{sec:sensitivity}. At each iteration, a gradient-based update is first applied to minimize the PAPR objective. The resulting intermediate solution is then refined through a sensing-aware stage that exploits the phase sensitivity of the AFDM autocorrelation. We adopt a derivative-free randomized local search that explores nearby solutions in the reserved-chirp subspace and retains the perturbation that yields the lowest ISL in the delay LAZ, \textit{i.e.}, $0<l \leq l_\text{max}$, denoted by $J_\text{LAZ}(\boldsymbol{r})$.

\subsubsection*{Randomized Local-search}\label{sec:ISL_step}
Starting from the intermediate solution produced by the PAPR-reduction step, a set of neighboring candidates is generated by applying small random perturbations $\boldsymbol r_{q,+}$ and $\boldsymbol r_{q,-}$ to the reserved vector,
\begin{align}
\boldsymbol r_{q,+}=\tilde{\boldsymbol r}^{(k+1)}+\delta\boldsymbol u, \, \text{and}, \,\boldsymbol r_{q,-}=\tilde{\boldsymbol r}^{(k+1)}-\delta\boldsymbol u,
\label{eq:candidates}
\end{align}
where $\boldsymbol u= \mathbf v_q / \|\mathbf v_q\|_2$, $\mathbf v_q\sim\mathcal{CN}(\mathbf 0,\mathbf I_{N_r})$. The procedure is repeated for a limited number of probes, which we verify in our experiments is generally small. The resulting AFDM waveform is reconstructed for each candidate, and the candidate yielding the lowest $J_{\rm LAZ}$ value is retained. This is equivalent to choosing the perturbation sign whose RCS-induced terms combine most destructively with the current sidelobe structure.
A complete summary of the projected gradient descent algorithm is presented in Algorithm 1.
\begin{algorithm}
\caption{Sensing-Aware CSR}
\label{alg:sensing-aware-csr}

\textbf{Input:} $\mathbf d$, $\mathbf A_{\mathcal R}^{H}$, $\alpha$, $\delta$, $p$, $K$, $K_{\rm p}$

\textbf{Initialize:} reserved chirp vector $\boldsymbol r^{(0)}$

\For{$k=0,\ldots,K-1$}{
Compute $\boldsymbol x^{(k)}=\mathbf d+\mathbf A_{\mathcal R}^{H}\boldsymbol r^{(k)}$

Compute the PAPR-oriented gradient $\nabla_{\boldsymbol r} f(\boldsymbol r^{(k)})$

Perform PAPR update:
\[
\tilde{\boldsymbol r}^{(k+1)}
=
\boldsymbol r^{(k)}
-
\alpha
\nabla_{\boldsymbol r} f(\boldsymbol r^{(k)})
\]

Set $\boldsymbol r_{\rm best}=\tilde{\boldsymbol r}^{(k+1)}$

Compute $J_{\text{LAZ}}(\boldsymbol r_{\rm best})$

\For{$q=1,\ldots,K_{\rm p}$}{
Generate random direction $\boldsymbol u$ as described in Section~\ref{sec:ISL_step}.

Form candidates as in \eqref{eq:candidates}.

Evaluate $J_{\text{LAZ}}(\boldsymbol r_{q,+})$ and $J_{\text{LAZ}}(\boldsymbol r_{q,-})$

\If{$J_{\text{LAZ}}(\boldsymbol r_{q,+})<J_{\text{LAZ}}(\boldsymbol r_{\rm best})$}{
Set $\boldsymbol r_{\rm best}=\boldsymbol r_{q,+}$ and
$J_{\text{LAZ}}(\boldsymbol r_{\rm best})=J_{\text{LAZ}}(\boldsymbol r_{q,+})$
}
\ElseIf{$J_{\text{LAZ}}(\boldsymbol r_{q,-})<J_{\text{LAZ}}(\boldsymbol r_{\rm best})$}{
Set $\boldsymbol r_{\rm best}=\boldsymbol r_{q,-}$ and
$J_{\text{LAZ}}(\boldsymbol r_{\rm best})=J_{\text{LAZ}}(\boldsymbol r_{q,-})$
}
}

Set $\boldsymbol r^{(k+1)}=\boldsymbol r_{\rm best}$ and normalize.
}

\textbf{Output:} optimized reserved chirp vector $\boldsymbol r^{(K)}$
\end{algorithm}

\subsection{Complexity Analysis}
The computational complexity of the proposed sensing-aware CSR algorithm arises from the PAPR gradient update and the random perturbation search. The gradient computation requires matrix-vector multiplications with complexity $\mathcal{O}(NN_r)$ per iteration; moreover, the objective function is Lipschitz continuous, and fast convergence of the gradient-based PAPR reduction algorithm is guaranteed \cite{rexhepi2025tone}. While the delay-LAZ ISL reduction stage evaluates $K_{\rm p}$ candidate directions, leading to a complexity of $\mathcal{O}(K_{\rm p}l_\text{max} N)$. Therefore, for $K$ outer iterations, the overall complexity is $
\mathcal{O}\left(K\left(NN_r+K_{\rm p}l_\text{max} N\right)\right)$.

\section{Numerical results}

\subsection{Simulation Parameters}
The algorithm outlined above was tested on an AFDM system employing 8-PSK. Unless otherwise stated, all simulations are conducted with \(N=128\) subcarriers and DAFT parameters \((c_1,c_2)=\left(\frac{21}{2N},\frac{1}{N^2}\right)\) as in \cite{meng2026afdm_papr_isl}. The maximum considered lag is \(l_{\max}=8\). The proposed optimization uses a gradient step size of \(\alpha=1\), a total of \(K=500\) iterations, \(K_p=2\) perturbation candidates per iteration, a \(p\)-norm approximation parameter of \(p=100\), and an ISL-control coefficient of \(\delta=0.3\). The initial RCS can be chosen randomly from the unit circle, at random positions in the AFDM sequence.
We examine the behavior of the proposed AFDM designs in terms of ISL, PAPR, and target detection, while comparing with classic and recent methods.
\subsection{PAPR reduction Evaluation}

\begin{figure}[t]
  \centering
  \includegraphics[width=1\linewidth]{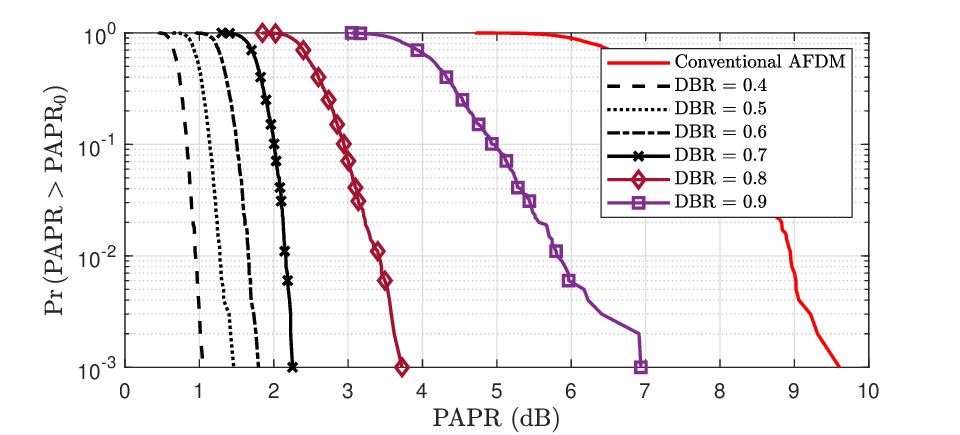}
  \caption{\centering The CCDF of the reduced PAPR with different $\rm DBR$ values.}
\label{schema}\label{fig:fig1}
\end{figure}

We first investigate the PAPR reduction performance without applying the ISL reduction step. Fig.~\ref{fig:fig1} shows the complementary cumulative distribution function (CCDF) of the resulting PAPR \textit{i.e.}, $\Pr\left(\mathrm{PAPR}>\mathrm{PAPR}_0\right)$, which represents the probability that the PAPR exceeds a certain threshold $\mathrm{PAPR}_0$, for different $\rm DBR$ values within the limits of desired spectral efficiency of communication systems. As expected, higher $\rm DBR$ values lead to smaller PAPR reductions, since the CSR contains fewer reserved resources (or degrees of freedom) available for optimization. Nevertheless, a significant PAPR reduction is achieved even when only 20\% of the resources are reserved. In particular, a reduction of 5.87 dB is obtained at $10^{-3}$ of the CCDF.

\begin{figure}[t]
  \centering
  \includegraphics[width=1\linewidth]{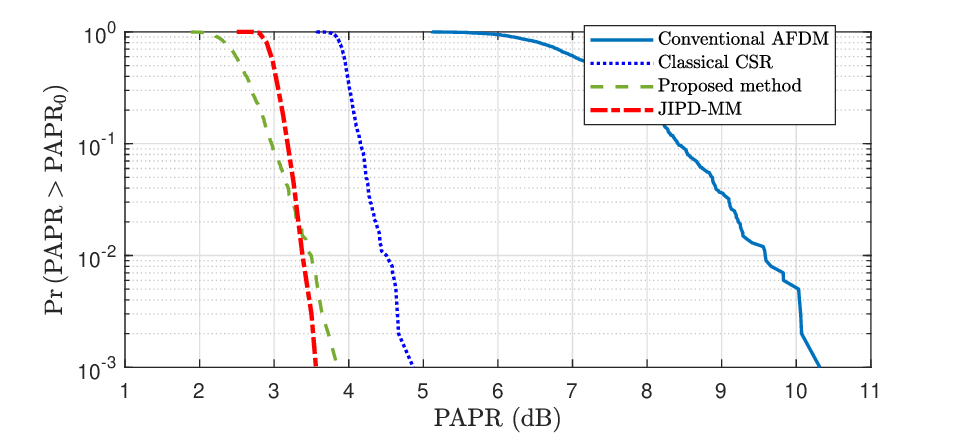}
  \caption{\centering The CCDF of the reduced PAPR using different techniques with $\rm DBR$ = 0.8.}
\label{schema}\label{fig:fig4}
\end{figure}

Next, we compare the proposed low-complexity PAPR reduction technique with other CSR-based approaches in Fig.~\ref{fig:fig4}. First, we consider the classical CSR method, which can be viewed as a straightforward extension of the OFDM tone reservation (TR) technique to the $\rm DAFT$ domain, where only the modulation operation is modified. Therefore, the parameter settings used for the proposed CSR scheme are also applied to the classical CSR method. The target PAPR of the latter is set at 4 dB, and the number of iterations is fixed at 30, to account for conventional usage parameters. We additionally compare our approach with the method proposed in \cite{meng2026afdm_papr_isl}, using its PAPR reduction mode based on RCS and pre-chirp parameters selection, called "ISL-PAPR-discrete-phase
majorization-minimization" (\textit{JIPD-MM}), which will be used throughout this section. The $\rm DBR$ is fixed at 0.8 for this experiment, and $N$ =128, meaning the RCS are of length $N_r$ = 26.
At CCDF = $10^\text{-3}$, the classical CSR baseline yields a PAPR of approximately 4.6 dB, while the JIPD-MM lowers it to 3.57 dB, and the proposed gradient-based CSR reduces it to approximately 3.85 dB.

\subsection{Sensing-Aware PAPR reduction Evaluation}
\begin{figure}[t]
  \centering
  \includegraphics[width=1\linewidth]{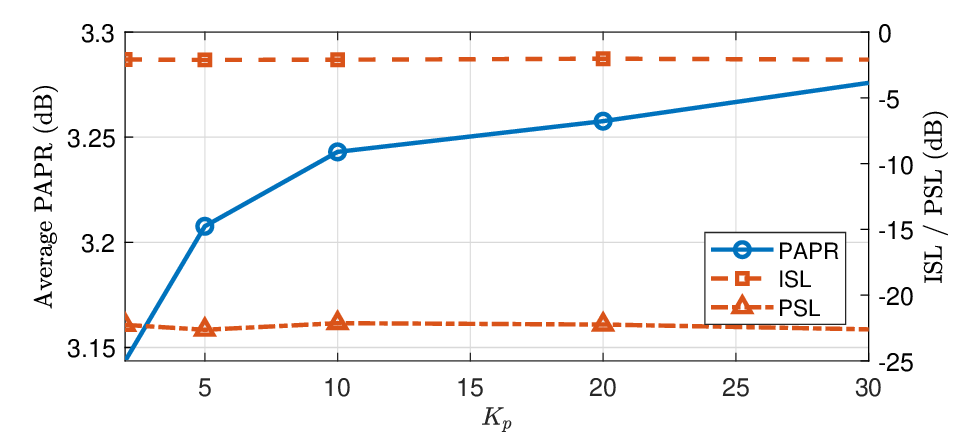}
  \caption{\centering Normalized ISL, PSL and PAPR versus $K_p$, $\rm DBR$=0.8.}
\label{schema}\label{fig:probes}
\end{figure}

Fig.~\ref{fig:probes} shows the LAZ ISL and the maximum sidelobe magnitude \textit{i.e.} the Peak-Sidelobe level (PSL), normalized to the mainlobe energy and the mainlobe magnitude, respectively, as well as the average PAPR versus the number of trials $K_{\rm p}$, with $\mathrm{DBR}=0.8$. Both metrics are nearly insensitive to $K_{\rm p}$, suggesting that a small number of trials is sufficient. This was expected from the small $\delta \in [0,1]$, which confines the search to a local neighborhood.

\begin{figure}[t]
  \centering
  \includegraphics[width=1\linewidth]{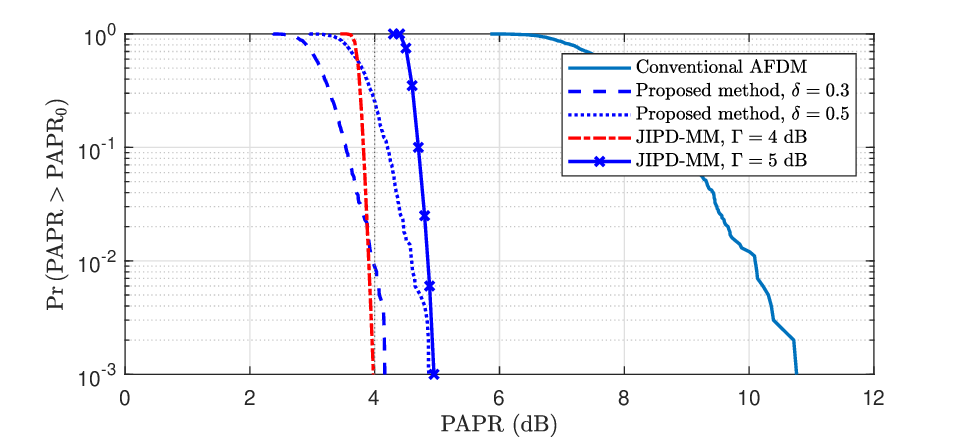}
  \caption{\centering The CCDF of the reduced PAPR using sensing-aware techniques, with $\rm DBR$ = 0.8. $\Gamma$ is the target PAPR in \cite{meng2026afdm_papr_isl}.}
\label{schema}\label{fig:fig7}
\end{figure}

Next, we investigate the behavior of the proposed sensing-aware method in terms of PAPR and ISL. Fig.~\ref{fig:fig7} shows the CCDF of the PAPR achieved using the proposed sensing-aware technique. As expected, introducing the ISL reduction stage results in a moderate increase in PAPR; moreover, the larger the ISL step size $\delta$, the greater the degradation in PAPR reduction.

\begin{figure}[t]
  \centering
  \includegraphics[width=1\linewidth]{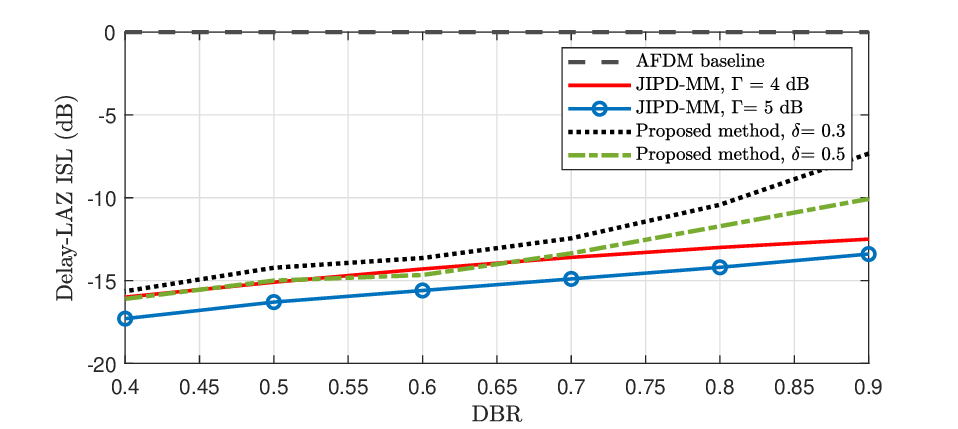}
  \caption{\centering Average ISL in the LAZ versus $\rm DBR$.}
\label{schema}\label{fig:fig6}
\end{figure}

\begin{figure}[t]
  \centering
  \includegraphics[width=1\linewidth]{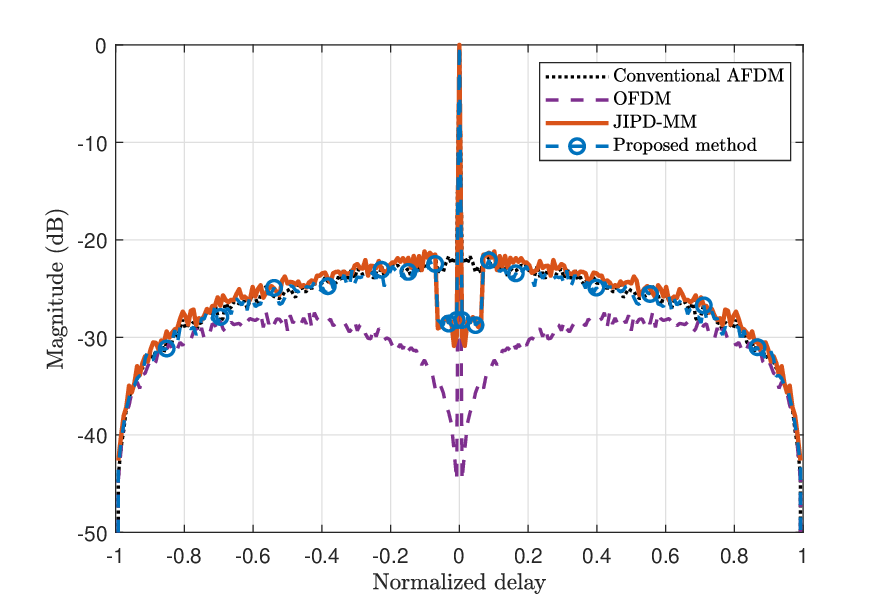}
  \caption{\centering ACF of the optimized AFDM waveforms with $\rm DBR$ = 0.8.}
\label{schema}\label{fig:fig8}
\end{figure}

To evaluate the ranging sidelobe suppression performance, Fig.~\ref{fig:fig6} depicts the average ISL within the delay LAZ as a function of the $\rm DBR$. The average ISL is reported in dB and normalized with respect to the conventional AFDM baseline, such that the AFDM curve remains fixed at 0 dB. As the number of RCS increases, the average ISL decreases. Moreover, the ISL achieves at least 10 dB reduction even with high $\rm DBR$ values. This gain can also be illustrated in the normalized ACF example shown in Fig.~\ref{fig:fig8} for $\rm DBR$ = 0.8, where the proposed scheme achieves approximately 7 dB sidelobe reduction in the LAZ, whereas the waveform generated using only the gradient-descent-based PAPR reduction method exhibits essentially unchanged sidelobe levels, coinciding with the conventional AFDM ACF curve. Furthermore, the performance of the proposed method is comparable to the method in \cite{meng2026afdm_papr_isl}, while requiring fewer operations. In fact, the JIPD-MM algorithm has a per-iteration complexity of $\mathcal{O}\left((|\mathcal{A}|\log N + L_PN)N\right)$, where $|\mathcal{A}|$ and $L_P$ are the number of samples in the delay-Doppler LAZ and the PAPR oversampling factor, respectively. As summarized in Table~\ref{tab:tab2}, for $N=128$, the proposed method requires approximately $1.1904\times10^{4}$ and $5.376\times10^{3}$ operations per iteration for $N_r=0.6N$ and $N_r=0.2N$, respectively, compared with $2.03\times10^{5}$ operations for the JIPD-MM approach, while achieving comparable performances.


\begin{table*}[!t]
\centering
\caption{Per-iteration computational complexity comparison for $N=128$.}
\label{tab:complexity_comparison}
\renewcommand{\arraystretch}{1.25}
\begin{tabular}{l | c| c |c}
\hline
\textbf{Method} & \textbf{Per-iteration complexity} 
& \textbf{$\rm DBR=0.4$} & \textbf{$\rm DBR=0.8$} \\
\hline
Proposed method (sensing-aware CSR)
& $\mathcal{O}\!\left(NN_r+K_p\l_{\max}N)\right)$
& $1.1904\times10^{4}$
& $5.376\times10^{3}$ \\
\hline
JIPD-MM \cite{meng2026afdm_papr_isl}
& $\mathcal{O}\!\left((|\mathcal{A}|\log_2 N+L_PN)N\right)$
& $2.03\times10^{5}$
& $2.03\times10^{5}$ 
\label{tab:tab2}
\end{tabular}
\end{table*}

\subsection{Sensing Performance}
\begin{figure}[t]
  \centering
  \includegraphics[width=1\linewidth]{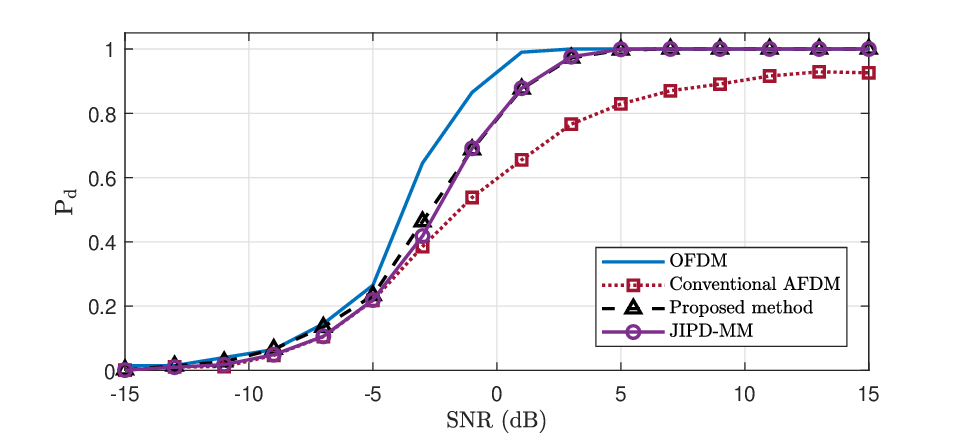}
  \caption{\centering Probability of detection vs SNR, using the optimized AFDM waveforms with $\rm DBR$ = 0.8.}
\label{schema}\label{fig:fig9}
\end{figure}

We evaluate the sensing performance of the optimized AFDM waveforms in terms of target detection. We consider a monostatic scenario where the transmitter and the receiver are colocated, with two same-speed closely-spaced targets (by 4 delay bins) to be detected, in which a strong target is 20 dB stronger than the weak target. Detection is performed by the Smallest-Of CFAR (SO-CFAR) detector on the range-Doppler map obtained with an AF-based method. We consider a transmit frame with 100 TD AFDM symbols. Fig.~\ref{fig:fig9} illustrates the weak-target detection rate $\rm P_d$ as a function of the Signal-to-Noise Ratio ($\rm SNR$) under false-alarm rate $\mathrm{P}fa$ = $10^{-4}$. Compared with the conventional AFDM baseline, the 7 dB sidelobe suppression (as shown in Fig.~\ref{fig:fig8}) effectively prevents the weak target from being masked over the entire $\rm SNR$ range. Moreover, at $\rm P_d = 0.8$, the proposed method provides a 5 dB SNR gain. We also observe that the sensing-aware technique has comparable performances and approaches the PSK-OFDM performance. Overall, despite \cite{meng2026afdm_papr_isl} considering an ISL reduction within the 2D delay-Doppler LAZ, the proposed method achieves similar detection performance with a lower computational complexity.

\section{Conclusion}

A sensing-aware CSR framework for AFDM has been proposed, combining gradient-based PAPR reduction with a randomized local-search stage for sidelobe suppression. By exploiting the sensitivity of the AFDM ACF to the RCS, the proposed method improves ranging performance without requiring full AF optimization. Numerical results demonstrate significant joint PAPR and ISL reduction, the latter translating into enhanced target detectability. In addition, the proposed approach achieves comparable sensing performance to the state-of-the-art while offering substantially lower computational complexity. A possible direction for future research is the reduction of the ISL over the entire delay-Doppler plane while maintaining the low computational complexity.

\bibliographystyle{IEEEtran}
\bibliography{main.bib}

@article{liu2022survey,
  title={{A survey on fundamental limits of integrated sensing and communication}},
  author={Liu, An and Huang, Zhe and Li, Min and Wan, Yubo and Li, Wenrui and Han, Tony Xiao and Liu, Chenchen and Du, Rui and Tan, Danny Kai Pin and Lu, Jianmin and others},
  journal={IEEE Communications Surveys \& Tutorials},
  volume={24},
  number={2},
  pages={994--1034},
  year={2022},
  publisher={IEEE}
}

@article{liu2025cp,
  title={CP-OFDM achieves the lowest average ranging sidelobe under QAM/PSK constellations},
  author={Liu, Fan and Zhang, Ying and Xiong, Yifeng and Li, Shuangyang and Yuan, Weijie and Gao, Feifei and Jin, Shi and Caire, Giuseppe},
  journal={IEEE Transactions on Information Theory},
  year={2025},
  publisher={IEEE}
}

@inproceedings{bemani2021afdm,
  title={AFDM: A full diversity next generation waveform for high mobility communications},
  author={Bemani, Ali and Ksairi, Nassar and Kountouris, Marios},
  booktitle={2021 IEEE International Conference on Communications Workshops (ICC Workshops)},
  pages={1--6},
  year={2021},
  organization={IEEE}
}

@article{rou2025normalized,
  title={Normalized Ambiguity Function Characteristics of OFDM, OTFS, AFDM, and CP-AFDM for ISAC},
  author={Rou, Hyeon Seok and de Abreu, Giuseppe Thadeu Freitas},
  journal={arXiv preprint arXiv:2510.11216},
  year={2025}
}

@article{meng2026afdm_papr_isl,
  title={A Unified Framework for Ambiguity Function Shaping and PAPR Control in AFDM Systems},
  author={Meng, Lingsheng and Guan, Yong Liang and Liu, Zilong and Luo, Yirui and Fan, Pingzhi},
  journal={arXiv preprint arXiv:2604.22198},
  year={2026}
}

@inproceedings{rexhepi2025tone,
  title={Tone reservation-based papr reduction using manifold optimization for ofdm-isac systems},
  author={Rexhepi, Getuar and Ranasinghe, Kuranage Roche Rayan and De Abreu, Giuseppe Thadeu Freitas and Gonz{\'a}lez, David},
  booktitle={2025 IEEE Wireless Communications and Networking Conference (WCNC)},
  pages={1--6},
  year={2025},
  organization={IEEE}
}

@article{vaezi2026tutorial,
  title={A tutorial on AI-empowered integrated sensing and communications},
  author={Vaezi, Mojtaba and Baduge, Gayan Aruma and Ollila, Esa and Vorobyov, Sergiy A},
  journal={IEEE Communications Surveys \& Tutorials},
  year={2026},
  publisher={IEEE}
}

@article{wang2006performance,
  title={Performance degradation of OFDM systems due to Doppler spreading},
  author={Wang, Tiejun and Proakis, John G and Masry, Elias and Zeidler, James R},
  journal={IEEE Transactions on wireless communications},
  volume={5},
  number={6},
  pages={1422--1432},
  year={2006},
  publisher={IEEE}
}

@article{zhang2026afdmenabled,
  author  = {Zhang, F. and Wang, Z. and Mao, T. and Jiao, T. and Zhuo, Y. and Wen, M. and Xiang, W. and Chen, S. and Karagiannidis, G. K.},
  title   = {{AFDM}-Enabled Integrated Sensing and Communication: Theoretical Framework and Pilot Design},
  journal = {IEEE Journal on Selected Areas in Communications},
  volume  = {44},
  pages   = {310--324},
  year    = {2026}
}

@article{tao2026affine,
  author  = {Tao, Y. and Wen, M. and others},
  title   = {Affine Frequency Division Multiple Access Based on {DAFT} Spreading for Next-Generation Wireless Networks},
  journal = {IEEE Transactions on Wireless Communications},
  volume  = {25},
  pages   = {4626--4641},
  year    = {2026}
}

@article{yuan2025papr,
  author  = {Yuan, H. and Xu, Y. and others},
  title   = {{PAPR} Reduction With Pre-Chirp Selection for Affine Frequency Division Multiplexing},
  journal = {IEEE Wireless Communications Letters},
  volume  = {14},
  number  = {3},
  pages   = {736--740},
  month   = mar,
  year    = {2025}
}

\end{document}